\documentclass[12pt]{article} \textheight=24 true cm
\usepackage{graphicx}
\usepackage{amsmath}
\begin{document}

\begin{center}
{\Large\bf Energy dissipation in wave propagation in general
relativistic plasma
}\\[8 mm]
Ajanta Das \footnote {Heritage Institute of Technology, Anandapur,
Kolkata -700107, India}
  and S. Chatterjee\footnote{Relativity and Cosmology Research Centre,
Jadavpur University, Kolkata - 700032, India, and also at IGNOU,
New Alipore College, Kolkata  700053,
e-mail : chat\_ sujit1@yahoo.com\\Correspondence to : S. Chatterjee} \\[6mm]

\end{center}
\begin{large}
\begin{abstract}
Based on a recent communication by the present authors the
question of energy dissipation in magneto hydrodynamical waves in
an inflating background in general relativity is examined. It is
found that the expanding background introduces a sort of dragging
force on the propagating wave such that unlike the Newtonnian case
energy gets dissipated as it progresses. This loss in energy
having no special relativistic analogue  is, however, not
mechanical in nature as in elastic wave. It is also found that the
energy loss is model dependent and also depends on the number of
dimensions.

\end{abstract}
\end{large}
   ~~~~~~~KEYWORDS : cosmology; higher dimensions; plasma

~~~PACS :   04.20, 04.50 +h
\section*{1. Introduction}
Interaction of electromagnetic waves with plasma, especially
reflection and attenuation of electromagnetic waves propagating in
plasma layers has attracted a great deal of attention in recent
years due to their potential applications in reflecting or
absorbing e.m wave energy, broadcasting signals in microwave
frequencies etc~\cite{china}. Moreover, radiation and energy loss
in cosmic plasma via Cerenkov radiation~\cite{li} as also Heavy
quark energy loss in a weakly coupled QCD plasma~\cite{cyrille}
were studied in the past. But here we discuss a situation where
energy dissipation in the propagation of an electromagnetic wave
through a cosmic plasma medium is not due to any scattering or
friction but due to the expanding background. In two recent
communications~\cite{dp1,dp2} we have investigated the propagation
of an electromagnetic wave through plasma in an expanding
background in the framework of higher dimensional spacetime(HD).
The present work is the continuation of the earlier two and
possibly the last one in the series where we specifically study
the question of energy dissipation of the electromagnetic wave in
its interaction with the plasma field. While great strides have
been made by general relativists to address the issues coming out
of the recent observations in the field of astrophysics and
cosmology and despite the fact that more than 90 percent of the
cosmic stuff in stellar interior and intergalactic spaces is made
up of matter in plasma state the much sought-after union between
the plasma dynamics and general relativity still remains elusive.
One inhibiting factor against the reunion is possibly the fact
that both Einstein's field equations and plasma equations are
highly nonlinear such that a combination of the two makes exact
analytical solutions very difficult to get forcing people the
alternative route to either numerical analysis or  a linearized
approximation of the plasma equations. However, following the well
known (3+1) formulation of general relativity by Arnowitt, Deser
and Misner(ADM) and its subsequent development and applications
for a covariant formulation of the equations of
magnetohydrodynamics(MHD)in general relativity by Thorne and
Macdonald~\cite{tm} there has been some spurt in activities in
general relativistic plasma~\cite{sc1,sc3,sc2,sc4}. While
classical MHD is rather well developed, not a great deal is known
about the GRMHD and partly because of the usual subtleties in
defining physically meaningful frame of reference in GR one must
be wary of applying the classical results with intense
gravitational fields. In his electrodynamics of moving bodies
Minkowski has given a covariant decomposition of the
electromagnetic fields, which was later extended by Pham Man
Quan~\cite{nan} and most extensively by Lichnerowicz~\cite{L} in
 GRMHD.In a series of articles listed earlier some of us studied
the interaction of electromagnetic waves with plasma, propagation
of Alfven wave and also found out the dispersion relations and
other related properties in an expanding background, for which as
a test case we have chosen spatially flat
Friedmann-Robertson-Walker (FRW) cosmology generalised to higher
dimensions.  This is the simplest background, yet it still
illustrates how the curvature as well as the nonstaticity of
spacetime can affect simple MHD results. For this simple metric it
is possible to split the ordinary electromagnetic field tensor,
$F^{\mu\nu}$ into the ordinary electric and magnetic fields E and
B in terms of which the field equations are more familiar. For our
present work the consequences of this simple decomposition is
nontrivial. This allows us to make use of the intuition from the
known results of flat MHD to be applied to its general
relativistic analogue. As the split formalism has been extensively
discussed in the literature we shall not , for brevity, restate
those things here but refer  the readers to reference [4].  We
have attempted the analysis in the framework of HD spacetime which
is an active area of activity in its attempt to unify gravity with
all other forces of nature. It also finds increasing applications
in brane inspired cosmology as also in STM theory~\cite{wesson}.
Moreover, it may also provide an alternative physical explanation
of the current accelerating era of the universe without bringing
in any hypothetical quintessencial type of scalar field
~\cite{sc5} by hand. For our case it assumes particular
significance because it is in the realm of early universe that
both plasma physics and HD are specially relevant. In fact it can
be shown that Einstein's equations generalized to higher
dimensions admit solutions where as the 3D space expands with time
the extra dimensions shrink in size as to be currently invisible
with the present day experimental technique. It is further
conjectured that some stabilising mechanism (quantum gravity may
be a possible candidate) stabilises the extra space to the
planckian scale. So the universe we see today appears manifestly
four dimensional.
\\ As mentioned earlier we have in the past rather
extensively discussed the propagation of electromagnetic waves
through different types of perfectly conducting plasma medium. For
the very simple, conformally metric form chosen for our test case
we find that most of the wellknown the general relativistic
results mimic the analogous newtonnian plasma mechanics, except
that all the field variables are now non static and share the
background expansion of the FRW metric. In this work we turn our
attention to another important aspect of wave propagation i.e.,
energy dissipation. In classical MHD one finds that no power is
dissipated during propagation for a perfectly conducting medium.
But the situation drastically changes for an expanding background
where energy dissipation does occur. It is observed that the
expanding background introduces a sort of dragging force to effect
the dissipation, which is not to be confused with the usual
mechanical drag. In the present work we have calculated the energy
dissipation using our solutions from the previous works. Although
we have mainly carried out the exercise in a higher
dimensional(HD) background we believe that most of our findings,
barring some qualitative differences, apply to the four
dimensions also.\\

\section*{2. Mathematical formalism}
\textbf{a. Newtonnian Mechanics}\\\\
Before embarking on the question of dissipation of electromagnetic
wave through a plasma medium in general relativity we try to
recapitulate, very briefly, the analogous situation in flat space.
From any standard textbook on Plasma physics we know that in
newtonnian mechanics the dielectric constant is hermitian which
implies no damping. This can be easily shown as follows: When
propagating through a plasma medium the electromagnetic variables,
for example, are given by
\begin{equation}
E = Re~ {E_{0}e^{i(k_{i}x - \omega t)}}
\end{equation}
\begin{equation}
J = Re~ {J_{0}e^{i(k_{i}x - \omega t)}}
\end{equation}

Here E and J may depend on $\omega$
 and k but not on time such that
the average power dissipation in a cycle is given by
\begin{equation}
P = E_{i}J_{i}= \frac{1}{4}[~ E_{0}J_{0}e^{2i\phi} +
E_{0}J_{0}^{*} + E_{0}^{*}J_{0}+ E_{0}^{*}J_{0}^{*}e^{-2i\phi}]
\end{equation}
where $\phi = k.x - \omega t$
 Again $ \textbf{J }= \sigma \textbf{E}$. Now for homogeneous
 spacetime the conductivity,  $\sigma$ behaves like a scalar. But
 for anisotropic spacetime (say an external magnetic field in a
 particular direction)it is, in general, a tensor of rank two
 because, as is wellknown, the linearized particle velocity
 \begin{equation}
 v = \frac{e}{i\omega m}( E + \frac{v}{c}\times B_{0})
\end{equation}
introduces an anisotropic  velocity field such that the
constitutive relation reduces to, $ J_{i}=\sigma_{{ij}} E_{j}$ and
we get for average dissipation over a complete cycle
\begin{equation}
< P > = \frac{1}{4} [~ E_{0i}\sigma^{*}_{ij}E^{*}_{0j}+
E_{0i}^{*}\sigma_{ij}E_{0j}]
 \end{equation}
 Now, for any arbitrary vectors and matrix $ A.M.B= B.M^{T}.A$
 where T refers to transposition.
 So the last equation implies that
\begin{equation}
 < P > = \frac{1}{4} E_{0i}^{*}((\sigma_{ij}^{* T} +
 \sigma_{ij})E_{0j}
\end{equation}
We also know from elementary plasma mechanics that the dielectric
tensor of the plasma medium  is related to the conductivity tensor
as
\begin{equation}
\epsilon_{ij} = \delta_{ij} +
\frac{1}{-i\omega\epsilon_{0}}\sigma_{ij}
\end{equation}
Now the dielectric tensor for cold plasma (anisotropic in general)
is hermitian, as can be checked in any standard text book which
necessitates that the conductivity tensor should be anti hermitian
i.e. $ \sigma_{ab}=- \sigma_{ab}^{*T}$. So there is no energy
dissipation in classical plasma dynamics when an electromagnetic
wave moves in a plasma medium with a static
background.\\\\\\

\textbf{b. General Relativistic Case}\\

The situation  changes drastically when a similar analysis is
carried out in a nonflat expanding background of arbitrary
dimensions. As mentioned in the introduction the authors of this
report studied [4, 5] the propagation of electromagnetic waves in
an expanding plasma background in the framework of Einstein's
field equations both in four and higher dimensions  taking
spatially flat Friedmann- Robertson -Walker metric for simplicity.
Using the well known 3+1 decomposition formalism of ADM  we get
the interesting results that the field variables mimic the
classical special relativistic results except that the sinusoidal
vibrations need to be replaced by Hankel functions and the field
parameters are no longer constants but share the background
velocity of the embedded metric. Secondly the close resemblance to
the special relativistic results may be due to the very simple
metrics form we have considered - the conformally flat FRW line
element. With more complicated background metric, we suppose, the
3+1 split would yield  significantly different results. We shall
not go into the details of our earlier works here  but to make the
present work more tractable and transparent we need to digress
time to time to one of our recent papers[4] very briefly and refer
to the relevant
equations only as and when absolutely necessary.\\
For our background space we take the (d+1) dimensional generalized
FRW space time as

\begin{equation}
ds^{2} = dt^{2} - A^{2} \left( dx^{2} + dy^{2} + dz^{2} +
d\psi_{n}^{2}\right)
\end{equation}
(n = 5, 6, 7,  ... , d ) \\where $A \equiv A (t)$ is the scale
function.\\ In an earlier work \cite{sc9} one of us extensively
discussed the (d+1) dimensional isotropic and homogeneous space
time and assuming an equation of state, $p = \gamma \rho$ found
the scale factor as ($p$ = pressure, $\rho$ = energy density)
\begin{equation}
A \sim t^{\frac{2}{d(1+\gamma)}}
\end{equation}\\We then wrote down the Maxwell's equations
(for more details see Mcdonald et al for 3+1 split) for this
metric. Before proceeding further let us ask the pertinent
question - why is it not possible to formulate the equations of
electrodynamics in manifestly covariant form using electric and
magnetic four vectors? This is at variance with the  case of
\emph{spin} in an external electromagnetic field where one can
\emph{define} a spin 4-vector $s^{\mu}$ whose spatial part reduces
to $s$ in the proper frame of the particle and so subject to the
constraint that $s^{\mu}u_{\mu}= 0$ with $u_{\mu}$ the 4-velocity
of it~\cite{jackson}. A possible answer to this question comes
immediately to mind- the electric and magnetic fields are not the
spatial component of any four vector. It is only a very particular
combination of their components which form a fully covariant
object, the electromagnetic field tensor, $F_{\mu\nu}$. Only using
this tensor the manifestly covariant form of the Maxwell's
equations can be achieved. However, if one chooses a preferred
coordinate system it is indeed possible to use the electric and
magnetic fields as (d+1)-vector (we are here considering a (d+1)
dimensional spacetime) and then finally write down the Maxwell's
equations in a fully covariant form~\cite{nunez}. As mentioned
earlier the present work investigates plasma physics in curved
spacetime. To make use of the intuition gained from the
conventional plasma dynamics it is preferable to split the
electromagentic tensor, $F_{\mu\nu}$ into electric and magnetic
fields E and B in terms of which equations are more familiar. This
requires choosing a particular set of fiducial with respect to
which E and B and other physical quantities are measured. In what
follows we shall presently see that for our simple background
metric  the electromagnetic field tensor via (3+1)decomposition
does decouple as d- dim. electric and magnetic field and for the
privileged fiducial observers(FIDOs) one may write\\
\begin{equation}
F^{\mu\nu} = E^{\nu}u^{\mu}-E^{\mu}u^{\nu}+ \varepsilon^{\mu\nu
\gamma\delta}u_{\gamma}B_{\delta}
\end{equation}
\begin{equation}
J^{\mu} = \rho_{e}u^{\mu} + j^{\mu}
\end{equation}

\begin{equation}
\Phi^{\mu}= \phi u^{\mu}+ A^{\mu}
\end{equation}
\begin{equation}
F^{*\mu\nu} = B^{\nu}u^{\mu}-B^{\mu}u^{\nu}+
\varepsilon^{\mu\nu}_{\gamma}E_{\gamma}
\end{equation}
where $u^{\mu}$ is the fiducial (d+1)-velocity and
$\epsilon_{\mu\nu\gamma\delta}$ is the (d+1) dimensional
Levi-Civita tensor.\\Here the RHS terms are measured by the
fiducial observers in the usual manner of flat spacetime, and
which therefore have the usual physical interpretations and are
orthogonal to $u^{\mu}$ whereas the LHS terms are the
reconstructed charge-current(n+1)- vector, $J^{\mu}$, (n+1)-vector
potential, $\Phi^{\mu}$ etc.\\Moreover the electric current,
$J^{\mu}$ is now the sum of the two terms corresponding to the
convection current and the conduction
current, $j^{\mu}$ respectively and $ j^{\mu}u_{\mu}= 0$\\
One can invert these relations to get
\begin{equation}
\rho_{e}= -J^{\mu}u_{\mu}
 \end{equation}
 \begin{equation}
 j^{\mu} =\gamma^{\mu\nu}J_{\nu}
 \end{equation}
 \begin{equation}
 E^{\mu} = F^{\mu\nu}u_{\nu}
 \end{equation}
 \begin{equation}
 B^{\mu} = -\frac{1}{2}\epsilon^{\mu\nu\gamma\delta}u_{\nu}F_{\gamma\delta}
 \end{equation}
 \begin{equation}
\phi = - \Phi^{\mu}u_{\mu}
\end{equation}
\begin{equation}
 A^{\mu}= \gamma^{\mu\nu}u_{\nu}
\end{equation}
We are now in a position to formulate the general relativistic
Maxwell's equations for our simple FRW metric to get(see ref.2)
\begin{eqnarray}
  \nabla.E &=&4\pi\rho_{e} \\
  \nabla.B &=& 0 \\
  \frac{\partial E}{\partial t} &=& KE + cA^{-1} \nabla \times B - 4\pi J \\
  \frac{\partial B}{\partial t} &=& KB - cA^{-1} \nabla \times E \\
  \frac{\partial \rho_{e}}{\partial t} &=& K \rho_{e} - \nabla .J ~ \textrm{(charge ~ conservation)}
\end{eqnarray}
and finally the particle equation of motion in $(d+1)$ dimensions
as
\begin{equation}
\frac{DA^{d-1}p}{D\tau} =  A^{d-1}q \left(E + A \frac{v}{c}\times
B \right)
\end{equation} or

\begin{equation}
\frac{Dp}{D\tau} = \frac{d-1}{d}Kp + q \left(E + A
\frac{v}{c}\times B \right)
\end{equation}

where
\begin{equation}
\frac{D}{D\tau} = \frac{1}{\alpha} \left(\partial_{t} + v.\nabla
\right)
\end{equation}
is the convective derivative and the d- momentum
\begin{equation}
p = m_{e}\Gamma v
\end{equation}
($m_{e} $ is the rest mass, $\Gamma$ is the boost factor, and v,
the d-velocity).

Here $\nabla.$ and $\nabla\times$ are the ordinary Minkowskian
divergence and curl in Cartesian co-ordinates. Thus we see that at
least for the very simple type of metric chosen the
electromagnetic field tensor of general relativity is split up and
gets decomposed as ordinary flat space electric and magnetic
field.\\In what follows we shall consider, for simplicity, the
small amplitude linear theory such that the convective derivative
simply reduces to ordinary derivative, $\frac{d}{dt}$.\\We see
that the equations(20, 21) have the form familiar from flat-
spacetime, Lorentz frame electrodynamics. They permit
one~\cite{hanni}  to charcterise E and B by electric and magnetic
field lines while the rest have a slightly different form with
some additional inputs from curved geometry(e.g., A and K
terms).\\In this section we investigate the situation where a
plasma in thermodynamic equilibrium is slightly disturbed through
the passage of an electromagnetic wave. We assume that an external
ambient magnetic field is also present. We, however, assume the
plasma medium to be cold so that the pressure can be neglected
when considering the particle equation of motion. In stellar
systems one often encounters situations where relaxation times are
much larger than the age of the universe so that collisions (hence
pressure) may be neglected. The effect of an electric field is not
generally seriously considered because of the well known Debye
shielding effect. The general problem of an electromagnetic wave
propagating along an arbitrary direction with the external
magnetic field is given by Appleton and Hartee in the Newtonian
case when studying the propagation of radio waves in ionosphere.
Considering the fact that a general solution with arbitrary
$\theta$ is very difficult to tackle in an expanding background
with arbitrary number of dimensions we shall restrict ourselves to
the cases when the electromagnetic wave propagates parallel and
perpendicular to the magnetic field. However the topic is of great
importance in astrophysics and space science where electromagnetic
wave propagation in magnetized plasma is very relevant.\\With the
set of equations split to (d + 1) formalism we are now in a
position to attempt applications in varied plasma phenomena.\\If
as usual we set $k_{i}c = \omega_{i}$ (the angular frequency of
the wave at some initial time $t = t_{i}$) then we get from the
above Maxwell's equations ( see reference 2 for details) \\
\begin{eqnarray}
    E^{\mu} &=& E_{0}^{\mu}i \sqrt{\frac{2k_{i}cd(1+\gamma)}{\pi\{d(1+\gamma)-
    2\}}}t^{-\frac{2}{1+\gamma}}e^{-i\omega_{i}t\frac{d(1+\gamma)}{d(1+\gamma)-2}
    t^{-\frac{2}{d(1+\gamma)}}}e^{ik_{i}.r}\nonumber \\
    &=& E_{0}^{\mu}i \sqrt{\frac{2k_{i}cd(1+\gamma)}{\pi\{d(1+\gamma)-2\}}}t^{-\frac{2}{1+\gamma}}e^{- i \frac{d(1+
    \gamma)}{d(1+\gamma)-2}\omega_{d}t}e^{ik_{i}.r}
\end{eqnarray}
where
\begin{equation}
    \omega_{d} = \omega_{i}t^{-\frac{2}{d(1+\gamma)}}
\end{equation}
gives a measure of the red shift of the photon due to background
expansion. For radiation dominated era $\gamma = \frac{1}{d}$ ,
$\omega_{d} = \omega_{i}t^{-\frac{2}{d+1}}$, so the rate at which
the frequency decreases is maximum in 4D universe. Moreover
damping is greater in radiation era.

This finding may have nontrivial implications for astrophysics. In
an earlier work \cite{sc7} one of us showed that the process of
nucleosynthesis in higher dimensional space time is markedly
different from that in 4D space time. So like the previous
classical case the equation (29) may again be written as

\begin{equation}
E^{\mu} = {E_{0}^{\mu}(x, t)e^{i(k_{i}x - \omega t)}}
\end{equation}
with the essential difference that here $E_{0}$ depends both on
space and time and all the other physical quantities like k and
$\omega$ share the expansion of the universe and
\begin{equation}
\omega = \omega_{i}\frac{d(1+\gamma)}{d(1+\gamma)-
2}t^{\frac{-2}{d(1+\gamma)}}
\end{equation}
Moreover, the exponent $(k_{i}x - \omega t)$ may be written in a
tensorial form as $ k^{a}x^{a}$, where the 4-vector $k^{a}$ =
$(\textbf{k}, -\omega t)$.

 Now referring again to our earlier paper we find that for a two
 component plasma
\begin{equation}
v=\frac{iqt^{\frac{2}{d(1+\gamma)}}}{m_{e} \Gamma \omega_{i}}E = -
\frac{ie}{m_{e} \Gamma \omega_{d}}E
\end{equation}
The last equation is very similar to the flat space case except
that here, $\omega_{d}$ is not a constant but shares the
background expansion. With $ J^{\mu} = n_{0}q v^{\mu}$ we get
\begin{equation}
J^{\mu} = J_{0}^{\mu}~ e^{i(k.r - \omega t)}
\end{equation}
 where
 \begin{equation}
J_{0} = -
\frac{e^{2}E_{0}}{m_{0}\omega_{i}}\sqrt{\frac{2k_{i}cd(1+\gamma)}
{\pi\{d(1+\gamma)-2\}}}t^{-\frac{2(1+d)}{d(1+\gamma}}
 \end{equation}
 This relation via equation(24) simplifies to
\begin{equation}
J = \frac{ie^{2}}{m_{e}\omega_{i}}t^\frac{-2}{d( 1+ \gamma)}~ E
\end{equation}
\\In a recent communication [4] we have shown that when an
 electromagnetic wave propagates through a plasma medium in a
 (d+1)dimensional expanding background in the presence of an ambient
 external magnetic field along the z direction the dielectric tensor comes out to be
 \begin{eqnarray}
  \epsilon_{11} &=& \epsilon_{22}=\epsilon_{44}=\epsilon_{55} =  ...
  = \epsilon_{dd}=1 - \frac{\omega_{p}^{2}}{\omega_{d}^{2}-\omega_{c}^{2}}= p_{1} \textrm{(say)}\\
  \epsilon_{12} &=& \epsilon_{14}=\epsilon_{15}= ... = \epsilon_{1d}
   =\frac{\omega_{c}}{\omega_{d}} \frac{\omega_{p}^{2}}{\omega_{d}^{2}-\omega_{c}^{2}} = p_{2}\\
  \epsilon_{31} &=& \epsilon_{32}= \epsilon_{34}= ... = \epsilon_{3d} = 0 \\
  \epsilon_{33} &=& 1 - \frac{\omega_{p}^{2}}{\omega_{d}^{2}} = p_{3}
\end{eqnarray} so that in matrix form
\begin{equation}
\epsilon_{\mu\nu} =\left(%
\begin{array}{cccccc}
  p_{1} & ip_{2} & 0& ip_{2} & . & ip_{d} \\
  -ip_{2} & p_{1} & 0 & ip_{2} & . & 0 \\
  0 & 0 & p_{3} & 0& . & 0 \\
  -ip_{2} & -ip_{2} & 0 & p_{1} &. & 0 \\
  . & . & . & . & . & . \\
  -ip_{d} & . & . &. & . & p_{1} \\
\end{array}%
\right)
\end{equation}
Here $\omega_{p}$ is the plasma frequency given by

\begin{equation}
\omega_{p}^{2} = \frac{b_{d} n_{0}e^{2}}{m_{e}}
  \end{equation}
  \begin{equation}
\omega_{c} = \frac{eB}{m_{e} c}t^{\frac{2}{d(1+\gamma)}}\equiv
\frac{e\hat{B}}{m_{e} c}
\end{equation}
where $\hat{B}$, the orthogonal magnitude of the ambient magnetic
field is given by $\hat{B}=|(B)_{z}(B)^{z}|^{1/2} = B
t^{\frac{2}{d(1+\gamma)}}$ for our system. Further the constant
\begin{equation}
b_{d}= \frac{2^{d/2}{\pi^{d/2}}}{(d-1)!!} ~~ (d-even)
\end{equation}
\begin{equation}
b_{d}= \frac{2^{(d +1)/2}{\pi^{(d -1)/2}}}{(d-2)!!}~~(d-odd)
\end{equation}
Thus the introduction of the magnetic field generates varied modes
transforming the dielectric constant scalar $\epsilon$ in equation
(35) to a second rank tensor $\epsilon_{ij}$. Although the
equations (30) - (40) exactly resemble the analogous expressions
in Newtonian theory the fact remains that all the frequencies now
depend on time rather than being constant. Further the cyclotron
frequency $\omega_{c}$ decays as $t^{-\frac{2(d-1)}{d(1+\gamma)}}$
exactly similar to the orthogonal component of the magnetic
field.\\A little introspection of the dielectric tensor shows that
like the special relativistic analogue it is the anisotropy
character of the plasma medium which introduces the tensorial
behaviour of conductivity because if the external magnetic field
is switched off , $p_{2}$ vanishes and $p_{1}=p_{3}$ and it
becomes a pure scalar. Moreover it is manifestly anti hermitian
when~$\omega$ is a real. So one would expect that like the
previous analysis the average energy dissipation will be nil. But
it is definitely not the case as the following analysis shows. Let
us once again recall the energy dissipation expression
\begin{equation}
P = {E_{\mu}J^{\mu}}= \frac{1}{4}[~E_{0\mu}J_{0}^{\mu}e^{2i\phi} +
E_{0\mu}J_{0}^{\mu*} + E_{0\mu}^{*}J_{0}^{\mu}+
E_{0\mu}^{*}J_{0}^{\mu *}e^{-2i\phi}]
\end{equation}
where $\phi = k.x - \omega t$ and $E_{01}~and ~J_{01}$ are both
functions of time due to the expansion of the background space
time. One may at this stage use the equation (7) to find that
\begin{equation}
\sigma_{\mu\nu} = \frac{-i\omega_{d}}{b_{d}}[\epsilon_{\mu\nu}-
\delta_{\mu\nu}]
\end{equation}
Following the previous argument and using the usual matrix
transformation rules  it follows that the second and the third
terms of the equation (46) give
\begin{equation}
\frac{1}{4}~E_{01}^{*}(\sigma_{ji}^{*}+\sigma_{ij})= 0
 \end{equation}
 This is like the special relativistic case discussed earlier but
 the similarity just ends there. Skipping intermediate steps we shall
 calculate the average value of the rest two terms over a complete
 cycle in some mathematical details to visualize the differences
 from the previous case. A long but otherwise straight forward
 calculation now gives that
\begin{equation}
<P> = \alpha (d)< t^{\frac{-2(1 + 2d)}{d(1+\gamma)}}~ 2Sin
2(kx-\omega t)>
\end{equation}
where $\alpha (d) = -\frac{E_{0}^{2}e^{2}d(1+\gamma)}{\pi m_{e}(1+
\gamma)d - 2}$ \\The equation(49) decomposes to
\begin{equation}
<P> =\alpha(d)~[~Sin~2k_{i}x\int_{0}^{T}
t^{\frac{-2(1+2d)}{d(1+\gamma)}}Cos~2\omega t dt -
Cos~2k_{i}x\int_{0}^{T} t
^{\frac{-2(1+2d)}{d(1+\gamma)}}Sin~2\omega t dt~ ]
\end{equation}
We have so far analyzed the whole situation  in (d+1) dimensional
spacetime with a general equation of state $p = \gamma\rho$. But,
as pointed out earlier, both higher dimensional cosmology and
plasma phenomena are most relevant in the very early phase of the
universe when the cosmology was in radiation dominated state with
$\gamma =1/d$. This input will considerably simplify the already
cumberous mathematical expressions although we believe most of our
inferences will be valid in a general equation of state also. For
the last term in equation(50) we find through equation (24)
\begin{equation}
\frac{1}{w_{i}}\beta^{\frac{4d}{d- 1}}\int_{0}^{T}
u^{\frac{-(7d+1)}{d- 1}}Sin~u^{2}du
\end{equation}\\
where $\beta t^{\frac{d - 1}{d + 1}}= u^{2}$ Here $ \beta =
\frac{2 m_{e}\pi\omega_{i}}{e^{2}E_{0}^{2}}~ \alpha_{d}$ . Also
the time period which also is sharing the expansion of the cosmos
is given by
\begin{equation}
T(t)= \frac{\pi}{\omega_{i}}~ t^{\frac{2}{(d + 1)}}
\end{equation}
The equation(51) integrates to
\begin{equation}
u^\frac{-(7d+1)}{d- 1}\int_{0}^{T}Sin~ u^{2}du  - \frac{-(7d -
1)}{(d-1)}\int_{0}^{T} u^\frac{-8d}{d-1}~ Sin~
[\frac{\pi}{\omega_{i}}~(\frac{u}{\sqrt\beta})^{\frac{4}{d-
1}}~]~du
\end{equation}
While dealing with the other expression we get almost similar
results. The special integrals we are dealing here are called
Fresnel Integrals which admit the following power series
expansions that converge for all $x$
\begin{equation}
S(x) = \int_{0}^{x} Sin~u^{2}du = \sum_{n=0}^{\infty}
(-1)^{n}\frac{x^{4n + 3}}{(4n+ 3)(2n+1)!}
\end{equation}
\begin{equation}
C(x) = \int_{0} ^{x}Cos~ u^{2}du = \sum
_{0}^{\infty}(-1)^{n}\frac{x^{4n + 1}}{(4n+ 1)(2n)!}
\end{equation}
The limits of these functions as $x$ goes to $\infty$ are known,
being equal to $ \sqrt\frac{\pi}{8}$. Also series expansion tells
of non zero value of it for any x. Thus we can conclude that the
average rate of doing work is non zero as an electromagnetic wave
passes through a plasma medium. This dissipation is unique and has
no special relativistic analogue coming as it does due to the
expansion of the background. As commented earlier that with
expansion  the background the density of the lines of force due to
the ambient magnetic field gets thinned out which, in turn,results
in the apparent damping of the magnetic field. It may  once again
be pointed out that this type of damping is not mechanical in
nature as one observes in Axionic dissipation or that due to
friction or viscosity or collision common in natural processes but
may be termed as expansion inspired damping.
\section*{3. Discussion :}
 We have here invoked the wellknown 3+1 split formalism of
physics to electrodynamics in FRW-like background. Although we
have worked out the problem of energy dissipation of a plasma wave
in HD spacetime we believe that most of the findings are valid, at
least qualitatively, in the 4D spacetime also. To start with the
FRW cosmology is chosen for the very simple reason that it is most
easy to handle, yet it illustrates how the curvature and
nonstaticity of the background can change the plasma MHD results.
In the process we have got the interesting result unlike the
special relativistic case where no dissipation occurs the
curvature and expansion do introduce a new phenomena of
dissipation unique in general relativity only without having any
classical analogue. This is not caused by the usual mechanical
friction-type forces as one frequently encounters in classical
mechanics. This may be interpreted as caused due to the thinning
out  of the magnetic field lines density due to background
expansion which results in the attenuation of the magnetic field
strength and power dissipation. As a future exercise one should
consider an inhomogeneous background to check what role
inhomogeneity plays in the whole process. Generalization to non
linear plasma may also introduce interesting physics.
\section*{Acknowledgement :} One of us(SC) acknowledges
the financial support of UGC, New Delhi for the award of a MRP.

\end{document}